\documentclass{aastex631}

\newcommand{\Msun}{\,{\rm M_\odot}}

\newcommand{\Mstar}{M_\star}

\usepackage{graphicx}
\usepackage[caption=false]{subfig}

\accepted{August 22, 2024}

\submitjournal{Research Notes of the AAS}

\begin{document}

\title{Sizes and Stellar Masses of the Little Red Dots Imply Immense Stellar Densities}

\correspondingauthor{Fabio Pacucci}
\email{fabio.pacucci@cfa.harvard.edu}

\author[0009-0009-7258-1637]{Carl Audric Guia}
\affiliation{Harvard College, Cambridge, MA 02138, USA}

\author[0000-0001-9879-7780]{Fabio Pacucci}
\affiliation{Center for Astrophysics $\vert$ Harvard \& Smithsonian, 60 Garden St, Cambridge, MA 02138, USA}
\affiliation{Black Hole Initiative, Harvard University, 20 Garden St, Cambridge, MA 02138, USA}

\author[0000-0002-8360-3880]{Dale D. Kocevski}
\affiliation{Department of Physics and Astronomy, Colby College, Waterville, ME 04901, USA}

\begin{abstract}
The ``Little Red Dots'' (LRDs) are red and compact galaxies detected in JWST deep fields, mainly in the redshift range $z=4-8$. Given their compactness and the inferred stellar masses in the hypothesis that LRDs are starburst galaxies, the implied stellar densities are immense. This Research Note uses an extensive catalog of LRDs from the PRIMER and the COSMOS-Web surveys to investigate these densities. We find a median (upper limit) on the effective radius of $80$ pc, which leads to median (lower limit) values of the core density of $\sim 10^4 \, \rm \Msun \, pc^{-3}$, and individual densities as high as $\sim 10^8 \, \rm \Msun \, pc^{-3}$, which is $\sim 10$ times higher than the density necessary for runaway collisions to take place. For $\sim 35\%$ of the LRDs investigated, the lower limits are higher than the highest stellar densities observed in any system in any redshift range.
\end{abstract}

\vspace{0.3cm}
\section{Introduction} 
\label{sec: intro}

The ``Little Red Dots'' (LRDs) are high-$z$ sources detected in several JWST deep fields (e.g., \citealt{Kocevski_2023, Harikane_2023, Maiolino_2023_new, Labbe_2023, Matthee_2023}), at a median redshift of $z\sim 6$ and with a distribution spread over $\sim 1$ Gyr of cosmic history \citep{Kocevski_2024, Akins_2024, Kokorev_2024_census}. These sources are strikingly red \citep{Matthee_2023} and extremely compact. While the median effective radius is $r_e \sim 150$ pc \citep{Baggen_2023}, some of them are significantly smaller, with $r_e < 35$ pc \citep{Furtak_2023_lensed}.

LRDs are either highly dense and compact starburst galaxies \citep{Labbe_2023, PG_2024}, or they host supermassive black holes (SMBHs) of $10^6 - 10^8 \Msun$ \citep{Harikane_2023, Maiolino_2023_new}, which are overmassive with respect to their host's stellar mass \citep{Pacucci_2023_JWST, Durodola_2024}; remarkably, despite showing broad emission lines typical of Type-1 AGN \citep{Greene_2023}, they are undetected in the X-rays \citep{Maiolino_2024_Xray, Yue_2024_Xray}, possibly because they are accreting at super-Eddington rates \citep{Pacucci_Narayan_2024}.

The stellar-only interpretation warrants further discussion. Stellar masses derived from SED fitting are large, in some cases $\sim 10^{11} \Msun$ \citep{Labbe_2023}. 
At high-$z$, the fact that the Balmer break redshifts out of the NIRCam bands may lead to overestimating such masses, alleviating the tension.
Measurements or, in most cases, upper limits on the effective radius are much firmer than any estimates on the black hole or stellar masses, and they are typically of the order of $\sim 100$ pc, which is $\sim 2\%$ of the effective radius of the Milky Way \citep{MW_size_2024}. Hence, the implied stellar densities are immense (as recently shown in \citealt{Baggen_2024_density} for three galaxies), and whether stable stellar systems can even exist in these conditions is unclear. 

In this Research Note, we consider the stellar mass and effective radius estimates from the Cosmos-Web and PRIMER surveys derived in \cite{Akins_2024} and \cite{Kocevski_2024} to infer the implied stellar densities.
\vspace{0.3cm}

\section{Methods} \label{sec:methods}

The catalog of LRDs provided by \cite{Akins_2024} contains 434 galaxies at $5<z<9$ from the COSMOS-Web survey.
This dataset was augmented with 64 PRIMER sources from \cite{Kocevski_2024}, removing a few duplicates based on their coordinates, thus leaving 475 sources for analysis.
We consider the inferred stellar masses in the star-only model and their (model-independent) effective radii to estimate the implied average stellar and core densities.

We compute an upper limit on the physical effective radius from the redshift of each source and its angular effective radius (measured in the F444W filter). Most of the LRDs in \cite{Kocevski_2024} and in \cite{Akins_2024} are unresolved.
The Plummer model \citep{Plummer_1911} is adopted to construct stellar density profiles and infer the core density of the LRDs; the associated potential is $\Phi(r) = -G\Mstar/\sqrt{r^2 + R^2_p}$, where $\Mstar$ is the total stellar mass, and $R_p$ is the Plummer scale radius. 
As is customary both in local and high-$z$ studies (see, e.g., \citealt{Baggen_2024_density}), we assume that the effective radius traces the half-mass radius and that $R_{\rm eff} = 1.3 R_p$.
The core density for each stellar system is calculated in the innermost $1$ pc radius. This choice is motivated by comparing our stellar densities with the maximum surface density reached by dense stellar systems, which is $\sim 10^{11} \rm \, \Msun \, kpc^{-2}$ \citep{Hopkins_2010, Akins_2024}, or $\sim 5 \times 10^4 \, \rm \Msun \, pc^{-3}$.

\begin{figure*}
    \centering
    \includegraphics[width=0.50\textwidth]{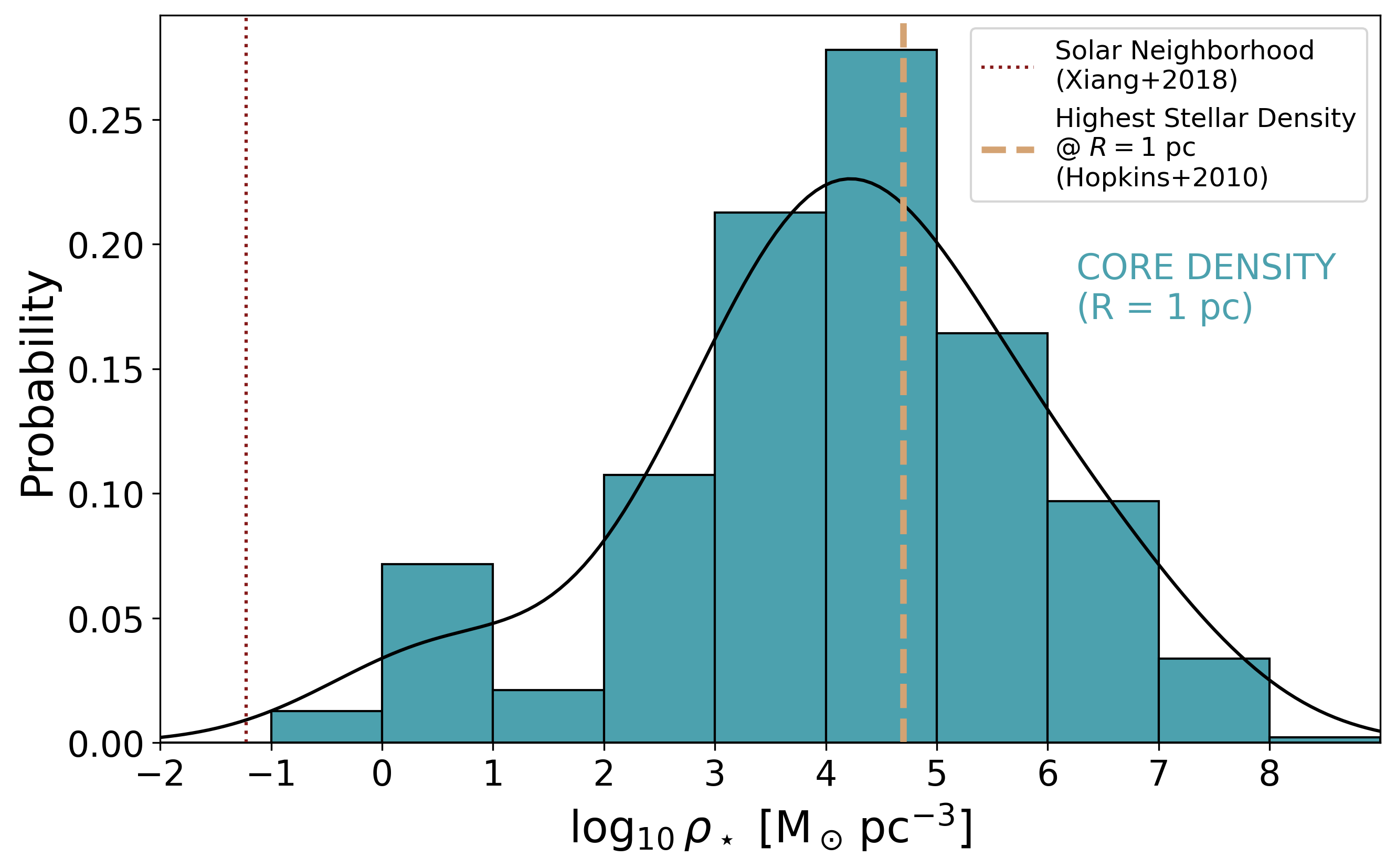} \hfill
    \includegraphics[width=0.49\textwidth]{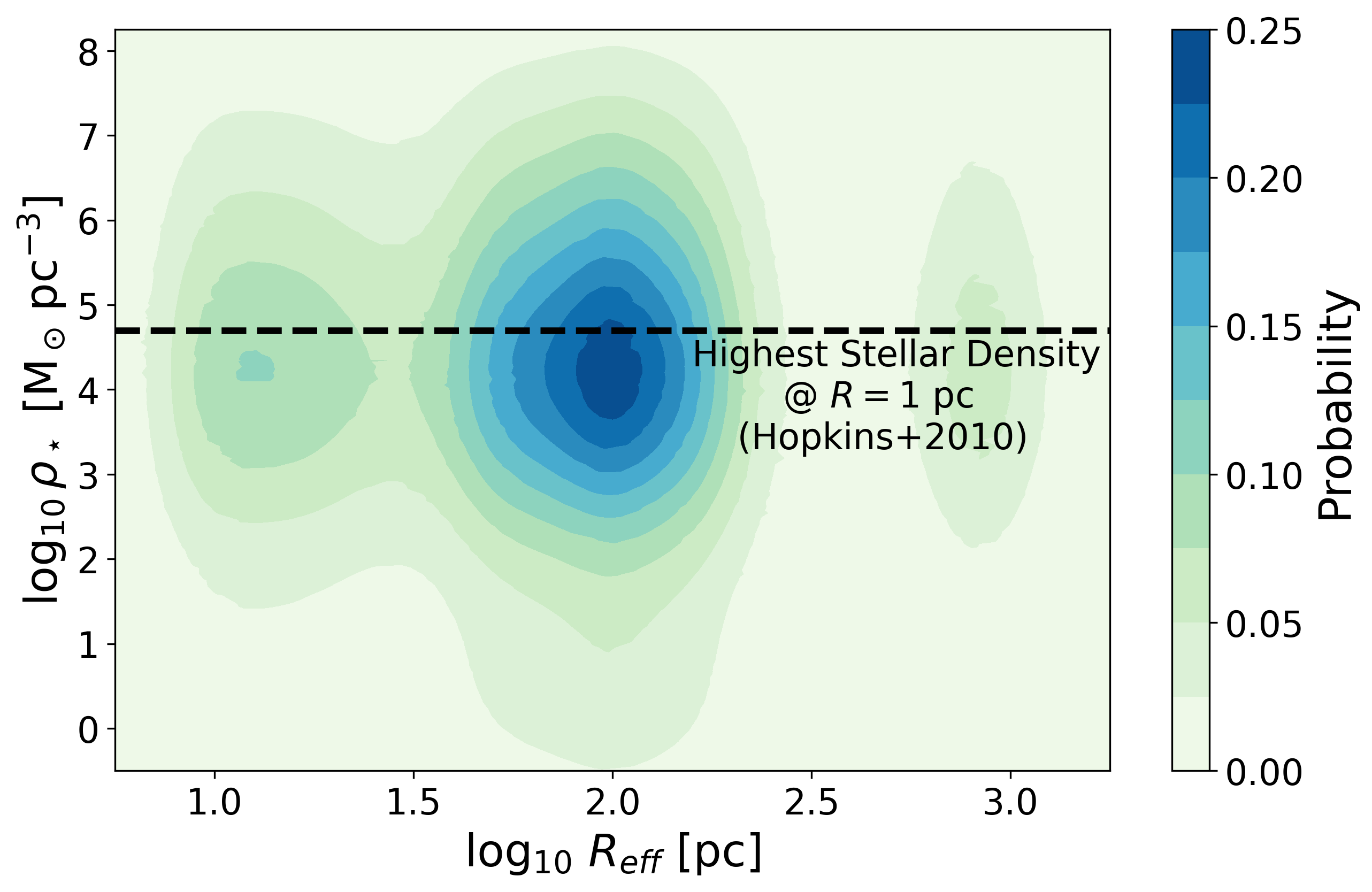}
    \caption{Statistical descriptors of the population of LRDs presented in \cite{Akins_2024} and \cite{Kocevski_2024}. \textbf{Left:} distribution of the inferred core stellar densities. \textbf{Right:} combined distribution of effective radii and densities.}
    \label{fig:densities}
\end{figure*}

\section{Results} \label{sec:results}
We display in Fig. \ref{fig:densities} (left) the distribution of the core density, computed with the mass within $1$ pc. 
All LRDs are characterized by implied core stellar densities that are significantly higher than the typical stellar density in the solar neighborhood (i.e., $\approx 0.06 \, \rm \Msun \, pc^{-3}$, \citealt{Xiang_2018}). 
Remarkably, most LRDs are significantly denser than typical Galactic globular clusters, characterized by core densities of $\sim 10^3 \, \rm \Msun pc^{-3}$ \citep{Djorgovski_1994}. In fact, the median value for the core density of the LRDs is $\sim 10^4 \, \rm \Msun pc^{-3}$. Interestingly, the maximum value reached in the core of any system is $\sim 10^8 \, \rm \Msun pc^{-3}$, which is $10$ times higher than the density necessary for runaway collisions to take place \citep{Ardi_2008, Fujii_2024}. 

To conclude, Fig. \ref{fig:densities} (right) displays the combined distribution of radii and core stellar densities for the LRDs. The most common LRD, expected to comprise around $24\%$ of the population, would be characterized by a core density of $10^{4.2} \, \rm M_\odot \, pc^{-3}$ with a radius of only $\sim 100$ pc. The distribution also demonstrates that LRDs as small as $10 - 20$ pc with core densities as high as $ \sim 10^{8} \, \rm \Msun \, pc^{-3}$ can be observed in around $5\%$ of the dataset.
The lower limits on the core stellar densities are, for $\sim 35\%$ of the LRDs, higher than the highest stellar densities observed in various stellar systems in any redshift range.

Since most effective radii measurements are upper limits, our density estimates are lower limits. Hence, it is crucial to investigate whether these extremely dense stellar systems are gravitationally stable.

\begin{acknowledgments}
    This project is funded by the Harvard College Research Program for Summer 2024. The Authors thank Chris Lintott for the helpful suggestions for improving the manuscript.
\end{acknowledgments}

\bibliography{ms}{}
\bibliographystyle{aasjournal}

\end{document}